# A new method of waveform digitization based on time-interleaved A/D conversion[*]

YE Chun-Feng (叶春逢)[1,2,3]   ZHAO Lei (赵雷)[1,2;1)]
FENG Chang-Qing (封常青)[1,2]   LIU Shu-Bin (刘树彬)[1,2]   AN Qi (安琪)[1,2]

[1] State Key Laboratory of Particle Detection & Electronics, University of Science and technology of China,
[2] Anhui Key Laboratory of Physical Electronics,
Department of Modern Physics, University of Science and technology of China, Hefei 230026, China
[3] 705 Research Division of Electronic Engineering Institute, Hefei 230037, China

**Abstract:** Time interleaved analog-to-digital conversion (TIADC) based on parallelism is an effective way to meet the requirement of the ultra-fast waveform digitizer beyond Gsps. Different methods to correct the mismatch errors among different analog-to-digital conversion channels have been developed previously. To overcome the speed limitation in hardware design and to implement the mismatch correction algorithm in real time, this paper proposes a fully parallel correction algorithm. A 12-bit 1-Gsps waveform digitizer with ENOB around 10.5 bit from 5 MHz to 200 MHz is implemented based on the real-time correction algorithm.

## 1  Introduction

Compared with the traditional method based on charging integration, shaping, and low speed data acquisition, modern waveform digitizers (WFDs) can be utilized to obtain the most comprehensive physics information in nuclear and particle physics experiments [1-9]. A number of waveform digitizing systems have been developed based on fast analog-to-digital converters (ADCs) or switched capacitor arrays (SCAs) [10, 11]. An ultra high sampling speed up to multiple Gsps can be achieved in SCAs; however, they have the disadvantage of a limited continuous recording length (e.g. DRS4, 1024-bin window) and a lower resolution (with an ENOB usually less than 8 bits). Therefore, great efforts have been devoted to the research of waveform digitization based on fast ADCs, which has been applied in physics experiments.

In the data acquisition system of the MAGIC telescope [7], for instance, relative slow flash ADCs (300-Msps 8-bit) are used as WFDs. Due to the limitation of the sampling rate, pulse stretching circuits are used to expand the input pulse to a FWHM of about 6.5 ns. Waveform digitization is also included in the readout electronics for LHAASO (the Large High Altitude Air Shower Observatory) KM2A [8], which is based on 12-bit ADCs with a sampling rate of 500 Msps. Considering that the rising time of photomultiplier tubes (PMTs) is about 3 ~ 4 ns，the sampling rate is not adequate to record the original waveform directly; thus a pulse stretching circuit is also needed. If a higher sampling speed is achieved, more detailed information of the original waveform can be obtained directly.

To meet the requirement of the ultra-fast WFDs beyond Gsps, the time interleaved analog-to-digital conversion (TIADC) method is employed in this paper. TIADC is an effective way to achieve a high sampling rate with relatively slower, cheaper and lower power consumption ADC chips, which is based on parallelism. In a TIADC system with *M* A/D conversion channels, the total sampling rate can be enhanced to *M* times higher by adjusting the sampling phases of the ADCs [12, 13]. However, due to the mismatches among different channels, additional errors are introduced. There exist three main mismatch errors, including the gain mismatch error, the offset mismatch error and the time-skew error between the clock signals distributed to them. These errors will ultimately limit the system performance, causing the so-called pattern noises and significantly degrading the signal to noise and distortion ratio (SINAD) [14-17]. Detailed discussions about the mismatch errors can be found in Refs. [18-21]. Different methods to correct these mismatch errors have been proposed. Correction of gain and offset mismatch errors is simple; however, correction of time-skew error is quite difficult, especially when it is implemented in real-time algorithm logic. There are several methods [22-28] for the time-skew error correction. Compared with the blind compensation [22, 23], interpolation [24] and fractional delay filters based method [25], the perfect reconstruction method [26-28] could be a favorable choice considering its simpler architecture and better feasibility in hardware implementation. Based on the perfection reconstruction, a multichannel-filtering approach was recently developed to further reduce computation complexity and the requirement on processing speed. However, under the situation of high resolution and high speed, it is still too difficult to implement the correction method in real-time algorithms.

[*] Supported by Knowledge Innovation Program of The Chinese Academy of Sciences (KJCX2-YW-N27) and National Natural Science Foundation of China (11175176, 10476028)
1) Email: zlei@ustc.edu.cn

This paper proposes a fully parallel correction algorithm that apparently has not been reported before, which overcomes the speed limitation in hardware design and can be utilized to correct the mismatch errors in real time. To further verify the efficiency of the method, a 12-bit 1-Gsps WFD is presented, which is composed of four 12-bit 250-Msps A/D conversion channels. The test results indicate that this WFD achieves an effective number of bits (ENOB) around 10.5 bits from 5 MHz to 200 MHz with real-time correction algorithms applied. This system also integrates a high performance interface based on Peripheral Component Interconnect (PCI) Express v2.0.

The outline of this paper is as follows: Section 2 describes the system architecture. Section 3 discusses a real-time correction method, which deliberates the parallel structure of the correction method and its implementation of the algorithms within a single FPGA device. In Section 4, we show a series of test results. In the last section, we conclude this paper and summarize what has been achieved.

## 2   System architecture

As shown in Fig. 1, the WFD consists of three main parts: front end circuits, multi-phase clock generation circuits, as well as real-time correction algorithms and the PCI Express interface integrated within one FPGA.

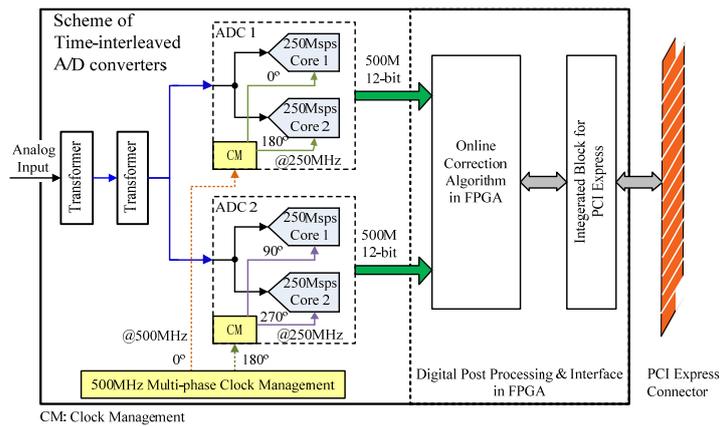

Fig. 1. (color online) Block diagram of the 12-bit 1-Gsps WFD.

The input analog signal is split into two channels, which is fed to two A/D converters KAD5512P50 [29]. Each ADC chip consists of two 12-bit 250-Msps A/D cores, the analog bandwidth of which is up to 1.3 GHz. There actually exist four A/D conversion channels in this system. A 500 MHz sampling clock is generated and converted to two signals with a phase shift of 180 degrees. Each of these two clock signals is inputted to one ADC chip, and further converted to a pair of 250 MHz signals with reverse phases, as the sampling clocks for two A/D cores. Thus, four interleaved sampling clocks with a phase shift of 90 degrees are finally obtained.

The output data streams of the four A/D cores are transferred to an FPGA device (XC6VLX130T). The data are processed with the real-time correction algorithms, buffered and then transferred to a personal computer (PC) through the PCI Express interface, which is also integrated in the FPGA.

**2.1 Front end circuits**

In high-speed high-resolution A/D conversion systems, the front end coupling circuit is a crucial part, which is used to implement single to differential conversion and impedance match. Cascaded transmission-line transformers (TC1-1-13M) are employed to provide additional isolation and improve common-mode rejection for a good balance between the differential output signals [30], as shown in Fig. 2.

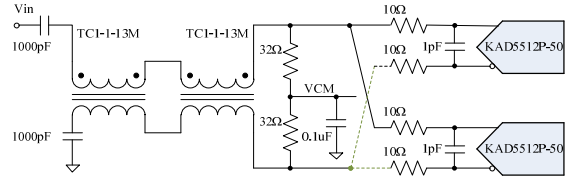

Fig. 2. (color online) Block diagram of the ADC front end circuits.

**2.2 Multi-phase clock generation circuits**

In clock generation circuits, there are two major issues with great concern -- the clock jitter and the time-skew between channels [31, 32]. Clock time-skew degrades the spurious free dynamic range (SFDR) and SINAD of an A/D conversion system and can be corrected by algorithms; meanwhile, the clock jitter deteriorates the signal-to-noise ratio (SNR) performance. The theoretical relationship [33] between clock jitter ($t_{jitter}$) and SNR is shown in

$$SNR = -20\log(2\pi f_{in} t_{jitter}) \quad . \quad (1)$$

Considering this is a 12-bit digitizer, we aim to achieve a SNR of 70 dB up to 200 MHz, which requires a maximum tolerated RMS jitter better than 251.6 fs. The structure of the clock generation circuits is shown in Fig. 3.

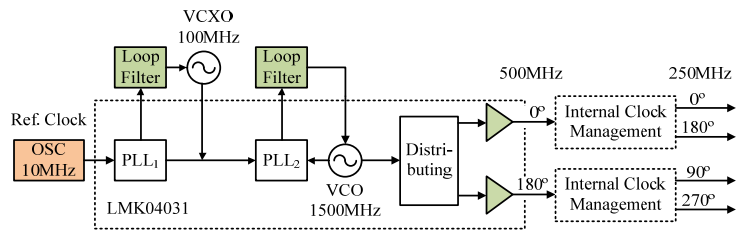

Fig. 3. (color online) Block diagram of the clock generation circuits.

A 10 MHz reference clock generated by a high performance oscillator VC-820 is fed to a low-noise clock jitter cleaner (LMK04031B) with cascaded phase-locked loops (PLL) inside. The reference clock is converted to a 100 MHz signal by the first PLL with an external voltage controlled crystal oscillator (VCXO) module (CVHD-950), which is sent to the second PLL with an internal voltage controlled oscillator (VCO) to generate a 1500 MHz signal. Through the divider in the distribution part in LMK04031B, two 500 MHz clock signals are obtained, which are based on the low voltage positive emitter-coupled logic (LVPECL) level. Ultra low jitter is achieved by allowing the external VCXO's phase noise to dominate the final output phase noise at low offset frequencies and the internal VCO's phase noise to dominate the final output phase noise at high offset frequencies. These result in best overall phase noise and jitter performance [34]. A phase shift of 180 degrees between the two 500 MHz clock signals can be achieved by reversing the positive and negative signal lines of the second clock. The internal clock management block inside the ADC chip further converts the 500 MHz clock signal to a pair of 250 MHz signals in phase opposition as the sampling clocks for two A/D cores. Thus, four interleaved sampling clocks in phase quadrature are finally obtained.

Simulation has been conducted to evaluate the phase noise of the clock circuits, as shown in Fig. 4. The integrated RMS jitter (100 Hz to 100 MHz) of the 500 MHz clock signal is estimated about 153.6 fs, which is beyond the requirement.

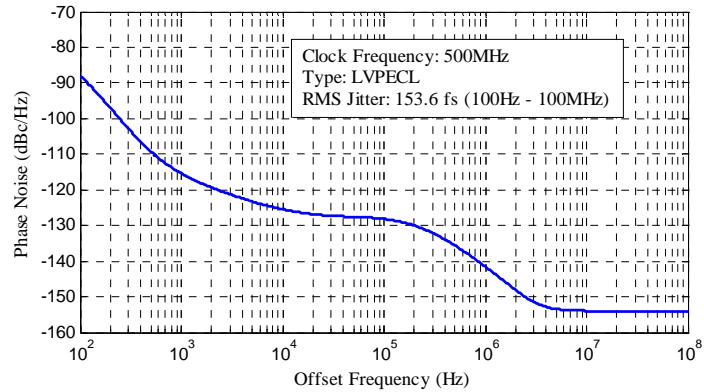

Fig. 4. (color online) Phase noise simulation results of the clock generation circuits.

**2.3  Multiple ADC synchronization**

The input 500 MHz clock signal is divided by two inside the ADC chip KAD5512P50, and then two 250MHz clock signals in phase opposition are supplied to the two ADC cores respectively. Fig. 5(a) shows a conceptual view of the internal data clocking management circuitry.

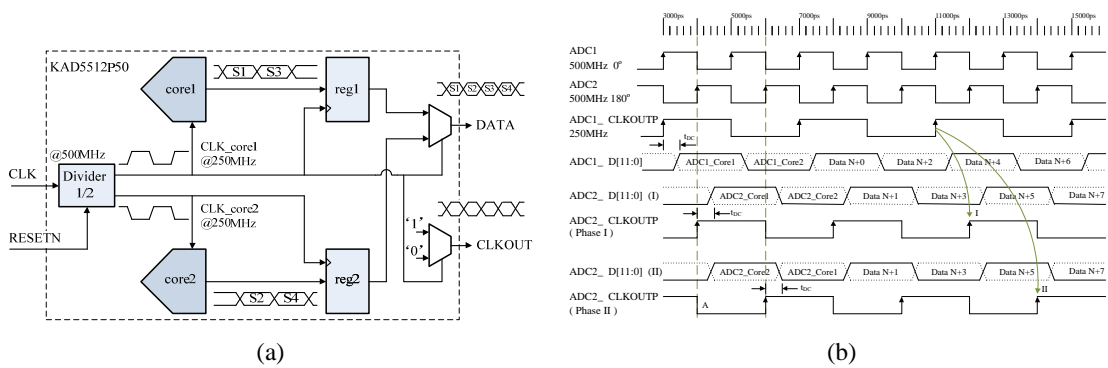

Fig. 5. (color online) Data clocking circuits inside KAD5512P50 and two possible sampling sequences among four ADC cores. (a) ADC internal data clocking management circuitry. (b) Two sample sequences among four ADC Cores.

This WFD consists of two ADC chips, each with two cores. The clock divider normally comes out of the reset signal in a random phase of the input clock [35]. So there exist two possible sampling sequences among the four ADC cores, as shown in Fig. 5(b). The first sampling sequence is ADC1_core1, ADC2_core1, ADC1_core2, ADC2_core2, when ADC2_CLKOUT (the output clock of ADC2) is in the state of Phase I; the second sampling sequence is ADC1_core1, ADC2_core2, ADC1_core2, ADC2_core1, when ADC2_CLKOUT is in the state of Phase II. Different filters are required to process the different sequences, so a predetermined sequence must be guaranteed in the circuit design.

As shown in Fig. 6, a reset circuitry is designed to generate two reset signals in our work, which are sent to the two ADC chips, respectively. These two reset signals are asserted high level in sequence in this design, and meanwhile a phase difference of 180 degrees between them is achieved (the phase difference between ADC1_CLK and ADC2 CLK is 180 degrees). Therefore, we can guarantee a 90 degrees phase difference among the four sampling clock signals of the ADC cores. With this sequence of the two reset signals, the sampling sequence is also determined -- the first sampling sequence in Fig. 5(b); therefore, the data streams of these two ADCs are well synchronized.

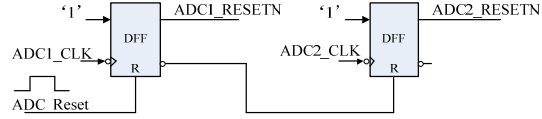

Fig. 6. (color online) Reset circuit for the two ADCs.

### 2.4 Real-time correction algorithms within the FPGA

As shown in Fig. 7, the 500-Mbps digital sequences of two A/D converters are received by a single FPGA with the double data rate (DDR) technique. By using the input serial-to-parallel logic resources (ISERDES) in the FPGA, the data streams are converted to four-channel data of 12-bit, 250 Msps, which are further deserialized to 16 channels with a data rate of 62.5 Msps. In the real-time correction block, the gain and offset errors are corrected by adders and multipliers, and time-skew error is corrected based on a fully parallel correction method, which will be presented in detail in Section 3.

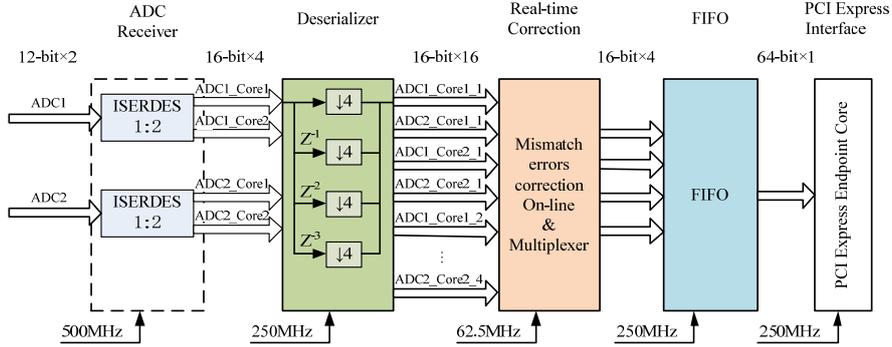

Fig. 7. (color online) Data processing procedure inside the FPGA.

## 3 Real-time correction methods

### 3.1 Algorithm architecture

As mentioned in Section 1, the perfect reconstruction method is a favorable choice for time-skew error correction. In this method, the filter banks are employed, where the ADC output data streams are processed by a digital FIR filter $F_i(z)$ with the impulse response $f_m[n]$ shown as below [36]:

$$f_m[n] = \begin{cases} \dfrac{M}{\pi} \cdot \dfrac{\prod_{i=0}^{M-1} \sin((n-d-d_i)\cdot\dfrac{\pi}{M})}{(n-d-d_m)\cdot \prod_{j=0, j\neq m}^{M-1} \sin((d_m-d_j)\cdot\dfrac{\pi}{M})} \cdot w_m(n), & |n-d-d_m| \leq L \\ 0, & |n-d-d_m| > L \end{cases} \quad (2)$$

Here, $d_m = t_m/T_s$, and $t_m$ denotes the time-skew between the other A/D channel and the first channel; $d$ denotes the delay which is a constant integer; $T_s$ denotes the period of the system sampling clock; $f_m[n]$ can be designed if the time skews $t_m$ are acquired. $w_m[n]$ corresponds to a Kaiser windowing function to smooth the time response of the filters. Those FIR filters are of the length $2L$.

However, this method requires up-sampling the data sequence of each A/D channel to the system sampling rate of $f_s$; therefore, it cannot be applied in real-time correction in ultra high speed situation. Poly-phase realization of the reconstruction filters was proposed [37], which increased the parallelism and reduced the computational complexity of the digital filter systems, since up-sampling was not further required. Thus, it is a feasible method to enhance the processing speed of the correction algorithm. The relation between the reconstruction filters $F_i(z)$ and the poly phase filters can be expressed as

$$F_m(z) = \sum_{k=0}^{M-1} z^{-k} F_{km}[z^M], \quad m = 0,1,...,M-1 \cdot \quad (3)$$

Here, $F_{km}(z)$ are the poly-phase components of $F_m(z)$, and the overall filter bank can be written as

$$F(z) = \begin{bmatrix} F_{0,0}(z) & F_{0,1}(z) & \cdots & F_{0,M-1}(z) \\ F_{1,0}(z) & F_{1,1}(z) & \cdots & F_{1,M-1}(z) \\ \vdots & \vdots & \ddots & \vdots \\ F_{M-1,0}(z) & F_{M-1,1}(z) & \cdots & F_{M-1,M-1}(z) \end{bmatrix}. \quad (4)$$

In this TIADC system, there are four A/D conversion channels of 250-Msps. With the poly-phase filter structure, the throughout processing rate is still too high to be applied in the logic design inside the FPGA. We employed a fully parallel filtering method, which reduces the requirement on the processing speed. Based on this method, we successfully implemented real-time correction algorithms in the FPGA.

Considering an $M$-channel TIADC system, as shown in Fig. 8, the output data stream of each ADC channel is converted to $N$-channel data sequences by a deserializer; therefore, a total of $M \times N$ channel data sequences are processed by the following poly-phase filter bank at a much lower speed than the original data rate.

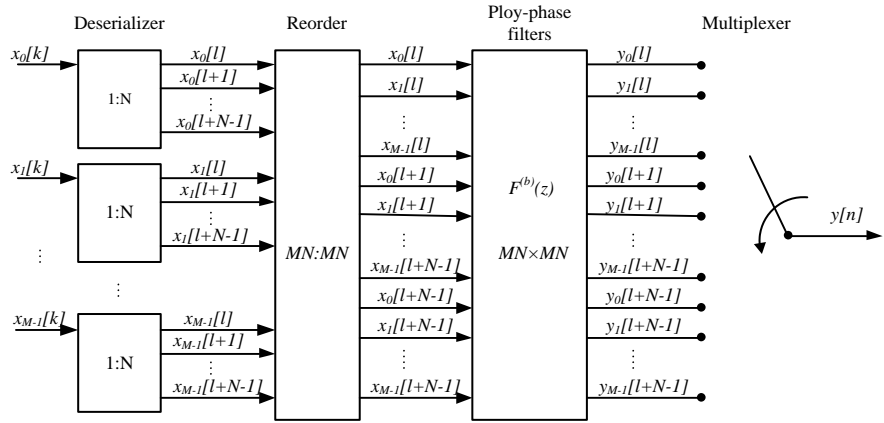

Fig. 8. Schematic of the fully parallel processing algorithm.

The kernel part is the poly-phase filter matrix of $MN \times MN$. The time-skew error among the $M \times N$ channel data sequences can be derived from that among the original M channel data streams. Based on the time-skew error values, the filter matrix can be implemented. As for this TIADC system, the filter matrix is of $16 \times 16$ with $N$ set to 4, and now the final processing speed is reduced to 62.5 Msps (250 Msps/4).

**3.2 Simulation of the fully parallel method**

In order to confirm the function of this fully parallel method, corresponding simulations have been conducted in a 4-channel TIADC system. The input signal $x_c(t)$ is supposed to be the sum of four sinusoid terms with a unit amplitude and frequencies of $f_s/15$, $2f_s/15$, $3f_s/15$ and $4f_s/15$ respectively. And we assume that the gain, offset and time-skew mismatches are [1, 1.02, 0.97, 1.03], [0, -2, 1, 3] $LSB$, and [0, -0.04, 0.02, -0.01] $T_s$. In the simulation, the continuous time signal $x_c(t)$ is quantized by the 12-bit 4-channel TIADC model, and the data streams of 4 channels are split into 16 channels (*N*=4). And now the time-skew mismatches are [0, -0.04, 0.02, -0.01, 0, -0.04, 0.02, -0.01, 0, -0.04, 0.02, -0.01, 0, -0.04, 0.02, -0.01] $T_s$. The reconstruction filter for each channel is designed to be of 80 taps, which is equivalent to 16 5-tap filter cells in one row of the matrix in Eq. (4).

Fig. 9(a) shows the frequency spectrum of $x_c(t)$ before correction, in which the distortions caused by the three types of mismatches can be observed. By applying the correction algorithms mentioned above, the mismatch errors can be effectively eliminated, as shown in Fig. 9(b).

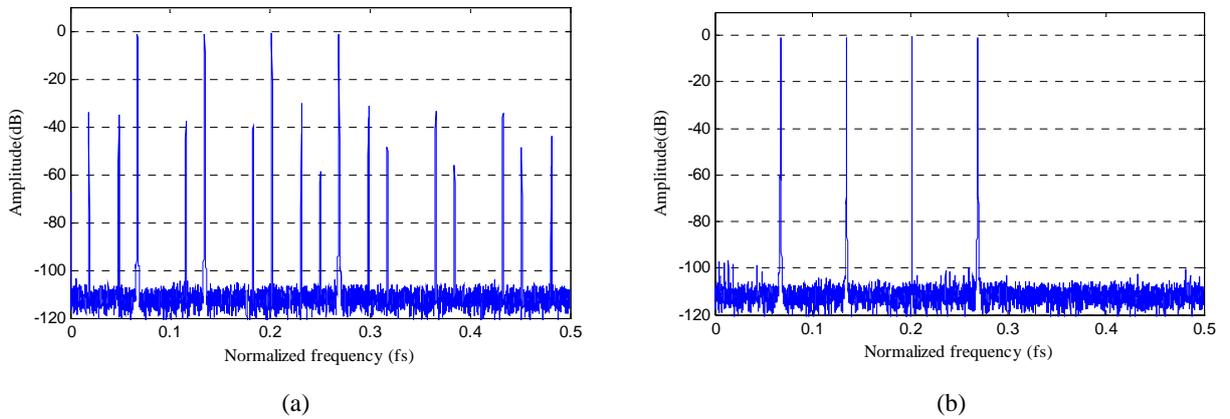

Fig. 9. (color online) Simulation results of the fully parallel filters. (a) The frequency spectrum before correction. (b) The frequency spectrum after correction.

**3.3 Real-time algorithm implemented in the FPGA**

We implemented the real-time algorithms within the FPGA device XC6VLX130T from Xilinx Virtex-6 family. The gain, offset, and time-skew mismatch errors are evaluated by the sine-wave fitting method [38]. As mentioned above, the correction algorithms for the gain and offset mismatch errors are comparatively simpler; therefore, we elaborate on the time-skew error correction. The time-skew error between the first ADC and the other three ADCs was measured, and the results is [0,-0.0076,-0.0046,-0.0089] $T_s$, with the maximum value around -0.00989 $T_s$.

Based on the aforementioned fully parallel method, the data stream of each A/D conversion channel is split into four channels; the four-channel TIADC system is converted to a 16-channel TIADC, with rearranged time-skew error of [0, -0.0076, -0.0046, -0.0089, 0,-0.0076, -0.0046, -0.0089, 0, -0.0076, -0.0046, -0.0089, 0, -0.0076, -0.0046, -0.0089] $T_s$. According to the simulation results in the Matlab, an 80-tap FIR filter for every channel is adequate for time-skew error correction. We can calculate the FIR filter coefficients according to Eq. (2), and further implement this 80-tap filter with 16 5-tap filter cells based on the poly-phase method, as shown in Fig. 10(a). Fig. 10(b) shows the structure of the 5-tap FIR filter matrix implemented in the FPGA.

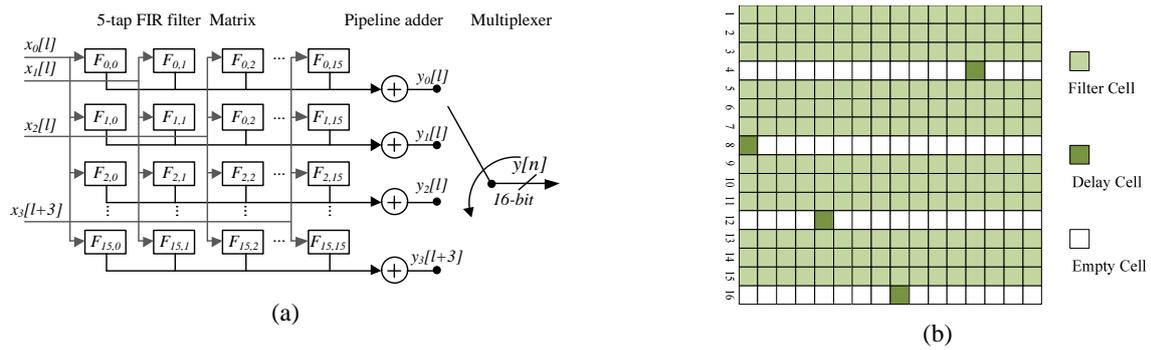

Fig. 10. (color online) Block diagram of the correction algorithms. (a) Structure of the poly-phase filter matrix. (b) Block diagram of the filter matrix implemented in the FPGA.

Based on the fully parallel filter structure, the processing data stream is decreased from 250 MHz to 62.5 MHz, which can be implemented easily in real time. As shown in Fig. 10(b), considering Channel 1, 5, 9 and 13 of the TIADC of the extracted 16 channels have no time-skew error, Row 4, 8 12 and 16 have just one delay cell in the poly-phase filter matrix and about 196 5-tap FIR filters are employed. In theory, a total of 980 multipliers are needed in this method. We verify this algorithm with a 12-bit 1-Gsps system, and of course we can implement these 980 multipliers operating at 62.5 Msps. If the speed is enhanced to 250 Msps, this fully parallel correction algorithm can support a real-time processing with a speed up to 4 Gsps. In this application, we multiplex the multiplications to reduce the logic resource consumption and about 392 multipliers (2 multipliers for one 5-tap FIR filter) are implemented, which operates at 187.5 MHz (3× 62.5MHz). The WFD photo is shown in Fig. 11.

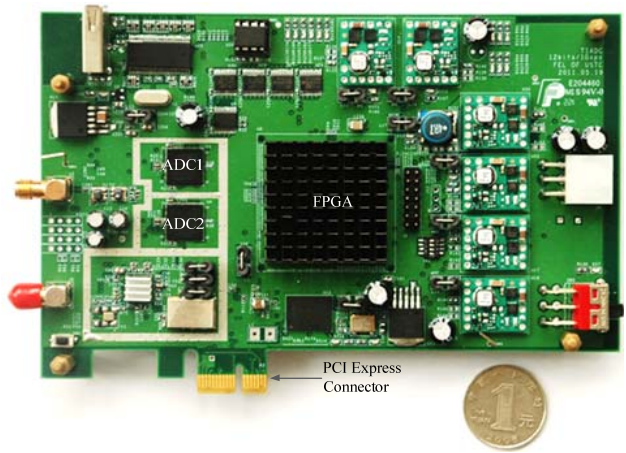

Fig. 11. (color online) The WFD photo.

## 4 Test results

To evaluate the system performance, a series of tests were conducted. The WFD is plugged into the PC mainboard via a standard PCI Express connector. The input signal from 5 MHz to 200 MHz is generated by a signal source R&S SMA100A, filtered by a coaxial band pass filter (BPF), and then input to the digitizer. The data results are transferred to the PC memory and analyzed with the software on the Matlab platform.

### 4.1 Mismatch error results

Tests were conducted with signals with different frequencies, and then the four parameter sine fit algorithm was applied to obtain the mismatch values, as shown in Fig. 12. Correction algorithms were then designed with these mismatch parameters.

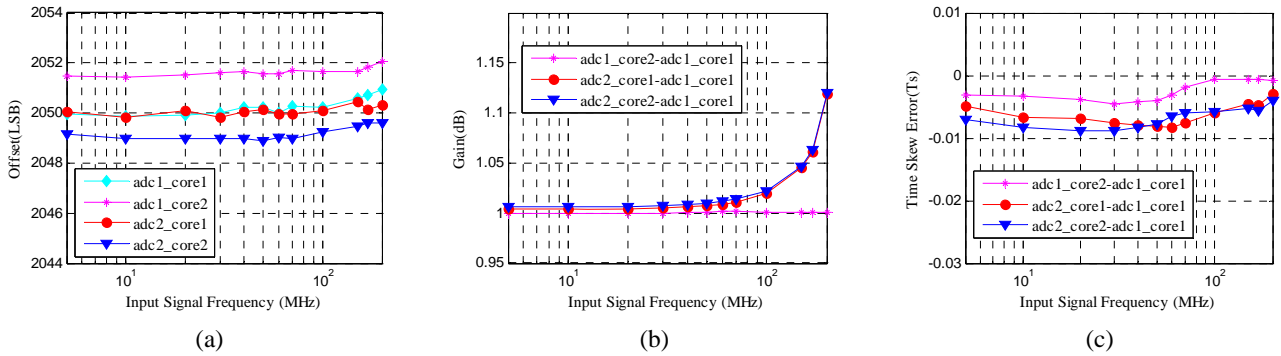

Fig. 12. (color online) Mismatch errors measured. (a) Offset mismatch error. (b) Gain mismatch error. (c) Time-skew error.

**4.2 Dynamic performance test results**

Dynamic analysis of this TIADC system is implemented based on the IEEE Std. 1241-2010 [38]. The data before and after correction are acquired, with a length of 32768 sample points in each channel (4 channels in totality). Analysis is performed through the Fast Fourier Transform (FFT) and spectral averaging method.

Fig. 14 shows the frequency spectra of test results with a 40.13 MHz input sinusoidal signal. Fig. 13(a) shows the frequency spectrum of the data from one single A/D conversion channel, no obvious spurious components are observed. By comparing Fig. 13(b) and Fig. 13(c) corresponding to the interleaved results before and after correction, the real-time correction algorithms are proven to function well.

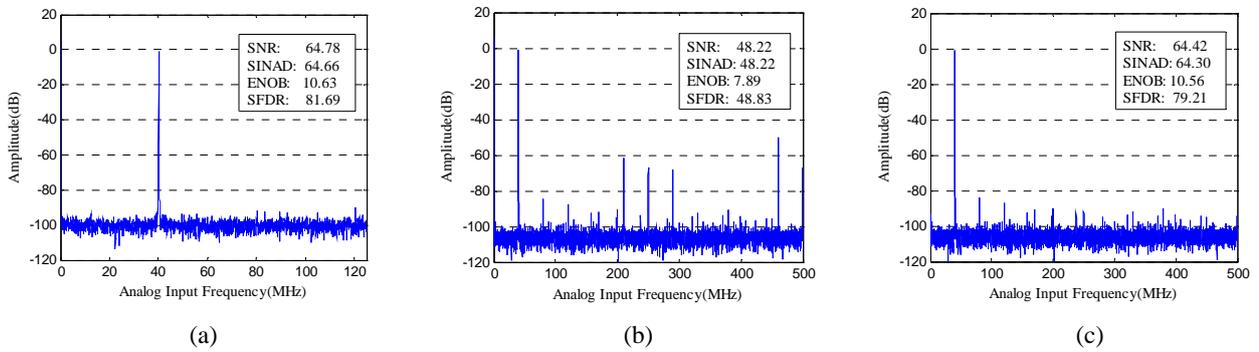

Fig. 13. (color online) The frequency spectra of test results with a input sinusoidal signal of 40.13 MHz. (a) The frequency spectrum of one single A/D converter channel with sample frequency of 250 MHz; (b) The frequency spectrum of the TIADC system with sample frequency of 1-Gsps before calibration; (c) The frequency spectrum of the TIADC system with sample frequency of 1-Gsps after calibration.

Turning the input frequency from 5 MHz to 200 MHz, systematic tests were conducted on SINAD, SFDR, ENOB and SNR, as shown in Fig. 14. The test results indicate that the real-time correction algorithms have significantly improved the system performance, achieving an ENOB around 10.5 bits from 5 MHz to 200 MHz.

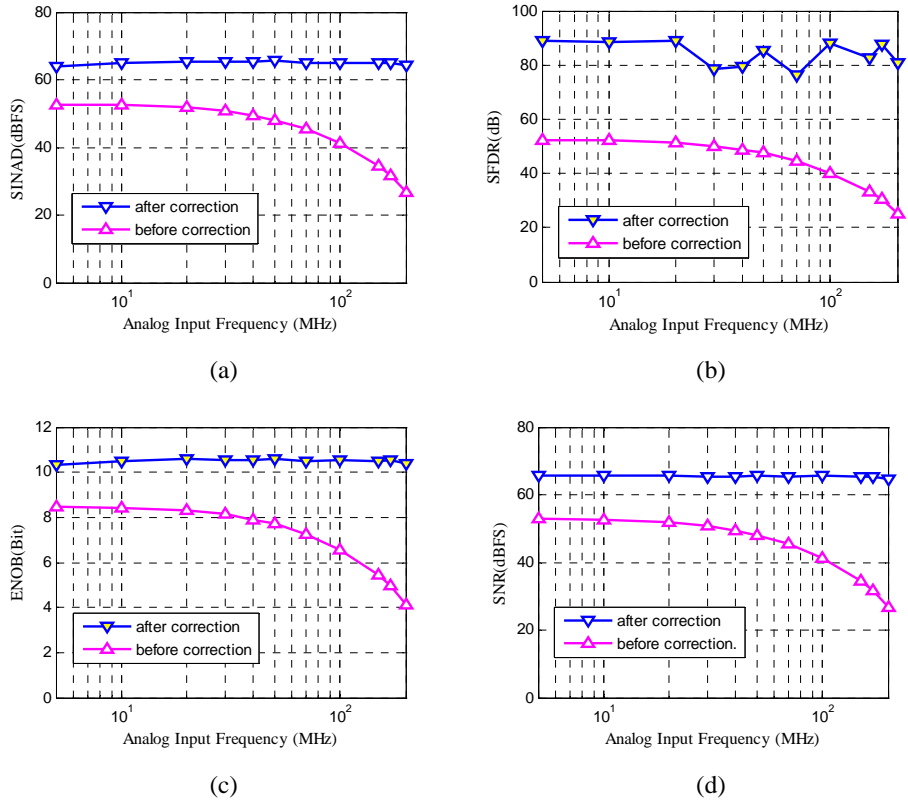

Fig. 14. (color online) Dynamic test results before and after correction. (a) SINAD test result. (b) SFDR test result. (c) ENOB test result. (d) SNR test result.

### 4.3 PMT pulse signal test result

PMTs are widely used in nuclear and particle physics [8, 40, 41]. We sampled a signal waveform from a PMT with a high-speed oscilloscope, and then used a 14-bit arbitrary waveform generator—Tektronix AFG3251 [42] to regenerate the PMT output signal (the rising time is 3.10 ns, the falling time is 9.90 ns) as the input pulse of this WFD. We also digitized this pulse with a LeCroy Oscilloscope 104MXi at the same sample rate of 1-Gsps for comparison. As shown in Fig. 15, no obvious difference exists between the results of the WFD and the commercial oscilloscope, except that the WFD exhibits a better performance with much lower noise.

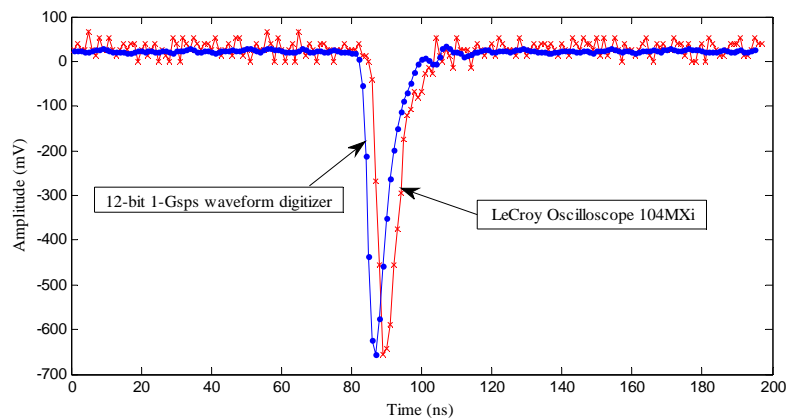

Fig. 15. (color online) Comparison of the PMT pulse test results between the WFD and a 1-Gsps oscilloscope.

## 5 Summary

In this paper we have proposed a fully parallel correction algorithm, which has been implemented further in an FPGA chip to correct the time-skew error among ADC channels of TIADC in real time. We have developed a 12-bit 1-Gsps WFD, by employing the fully parallel correction method in FPGA. Dynamic performance and PMT pulse tests were conducted, with the results indicating that this paper's methods do produce an efficient system performance, achieving an ENOB around 10.5 bits from 5 MHz to 200 MHz.

———————————————